\documentclass[journal=jacsat,manuscript=article]{achemso}

\usepackage[version=3]{mhchem} 
\usepackage{hyperref}
\hypersetup{
    colorlinks=true,
    linkcolor=[RGB]{255,0,0},
    citecolor=[RGB]{230,69,83},
}
\usepackage{xcolor}
\usepackage{soul}

\author{Jiayan Xu}
\affiliation{Department of Chemistry, Princeton University, Princeton, New Jersey 08544, United States}
\author{Abhirup Patra}
\affiliation{Shell International Exploration \& Production Inc., 200 N Dairy Ashford Rd, Houston, Texas 77079, United States}
\email{Abhirup.Patra@shell.com}
\author{Amar Deep Pathak}
\affiliation{Shell India Markets Pvt. Ltd. (Shell Projects \& Technology), Mahadeva Kodigehalli, Bengaluru, 562149, Karnataka, India}
\author{Sharan Shetty}
\affiliation{Shell India Markets Pvt. Ltd. (Shell Projects \& Technology), Mahadeva Kodigehalli, Bengaluru, 562149, Karnataka, India}
\author{Detlef Hohl}
\affiliation{Shell Information Technology International Inc., 3333 Highway 6 South, Houston, Texas 77082, United States}
\author{Roberto Car}
\affiliation{Department of Chemistry, Princeton University, Princeton, New Jersey 08544, United States}
\email{rcar@princeton.edu}

\title{Benchmarking Universal Machine Learning Interatomic Potentials for Supported Nanoparticles: Decoupling Energy Accuracy from Structural Exploration}

\begin{document}


\newpage
\begin{abstract}
    Supported nanoparticle catalysts are widely used in the chemical industry. 
    Computational modeling of supported nanoparticles based on density functional theory (DFT) often involves structural searches of stable local minimum energy configurations and molecular dynamics simulations at finite temperature.
    These are computationally demanding tasks that are intractable within DFT for large systems.
    In the last two decades, machine learning interatomic potentials (MLIPs) have been successfully used to substantially increase the size and time scales accessible to simulations approximating DFT accuracy.
    However, training reliable MLIPs is non-trivial as it requires many costly DFT calculations.
    Recently, several universal MLIPs (uMLIPs) have been developed, which are trained on large datasets that cover a wide range of molecules and materials.
    Here, we benchmark the accuracy and the efficiency of these uMLIPs in describing Cu nanoparticles supported on Al$_2$O$_3$ surfaces against our domain-specific DP-UniAlCu model.
    We find that the MACE-OMAT can reproduce reasonably well the low-energy structures found in global optimization at an energy accuracy comparable to DP-UniAlCu.
    Interestingly, the MatterSim-v1.0.0-1M model, which exhibits larger deviations in the binding energies, can find even more stable configurations than the other two models in some supported nanoparticle sizes, showing its capability in structure exploration.
    For MD simulations, MACE-OMAT and MatterSim-v1.0.0-1M can qualitatively reproduce the mean-squared displacements of Cu atoms (MSD$_\mathrm{Cu}$) predicted by DP-UniAlCu, albeit at roughly two orders of magnitude higher cost.
    We demonstrate that the uMLIPs can be very useful in simulating supported nanoparticles even without any fine-tuning, though their reduced efficiency remains a limiting factor for large-scale simulations.
\end{abstract}

\newpage

\section{Introduction}
Atomic simulation is an indispensable tool in understanding various physical phenomena from an atomic perspective.
Heterogeneous catalysis, a complex process that usually involves small molecules and solid surfaces, has benefited from valuable insights obtained from atomic simulation.\cite{Chen2020ChemRev}
The proper description of catalytic reactions and catalyst structures requires a reliable method at the quantum mechanical level.
Density functional theory (DFT) has been the workhorse for atomic simulation over the years.
However, due to its intensive computational cost, DFT-based simulations cannot satisfy the growing interest in the realistic modeling of catalytic systems that require both extended spatial and time scales.\cite{Grajciar2018ChemSocRev}

Supported nanoparticle catalyst is one of the most widely used catalysts in the chemical industry.\cite{Liu2018ChemRev}
The computational modeling of supported nanoparticles often involves the structure search of stable configurations at 0 K and the molecular dynamics (MD) simulation at finite temperature.
Both the structure search and the MD are computationally demanding and become intractable to DFT when the system size increases.
The typical size of a supported nanoparticle synthesized in experiments is in the range of 1-10 nm,{\cite{Dai2018ChemSocRev}} where a reasonable structure model of a 2-nm-sized nanoparticle together with the substrate may contain thousands of atoms.
The supported nanoparticle structure adopted in DFT-based studies is usually a simplified model that only contains a few dozen atoms.
Thus, the computational results are difficult to directly compare with experiments.
To understand and design better supported nanoparticle catalysts computationally, an accurate and efficient simulation method is highly desirable.

In the last two decades, machine learning interatomic potentials (MLIPs) have been developed to accelerate atomic simulations, which approximate DFT accuracy at almost force-field cost, and find extensive applications in heterogeneous catalysis.{\cite{Cheng2024PrecisChem,Omranpour2025ACSCatal,Xia2025ChemSocRev}}
However, the development of an MLIP is a non-trivial task, which requires the user to collect a large amount of \textit{ab initio} data through many training-validation-application cycles.
Also, the MLIP is usually specific to one system, and its transferability is limited.
For example, an MLIP trained on one crystalline phase may be useful for exploring configurations of another phase, but the energy and the forces are normally less accurate.
Recently, several universal MLIPs (uMLIPs) have emerged, namely, DPA2\cite{Zhang2024npjComputMater}, DPA3,\cite{Zhang2025arXiv} MACE-MP\cite{Batatia2025JCP}, and MatterSim\cite{Yang2024arXiv_mattersim}.
These uMLIPs are trained on large datasets that cover a wide range of molecules and materials, such as Materials Project (MP),\cite{Jain2013APLMater} Open Catalyst Project (OC2M),\cite{Tran2023ACSCatal} and Open Materials Dataset (OMAT),\cite{BarrosoLuque2024arXiv} and are expected to apply to a wide range of systems.
In general, uMLIPs provide a good foundation to develop system specific MLIPs because they can generate meaningful configurations without active learning and can be further fine-tuned to a target accuracy if necessary.{\cite{Liu2025arXiv}}
MACE-MP\cite{Batatia2025JCP} has been tested on numerous systems from solid materials to heterogeneous interfaces, while other recent uMLIPs have been mainly tested with solid-state systems.
Despite all the tests that have been done, the benchmark of the uMLIPs on supported nanoparticles is still lacking.

In this work, we examined several uMLIPs on simulating supported nanoparticles, where two main application scenarios are demonstrated: (1) global optimization that produces low-energy structures of nanoparticles serving good initial structures for the further investigation of dynamics, and (2) MD that reveals the dynamics of supported nanoparticles at finite temperature.
We choose Cu nanoparticles on Al$_2$O$_3$ surfaces as the benchmark system and set the domain-specific deep potential (DP) model developed in our previous work\cite{Xu2025ACSNano} as the benchmark baseline, which is denoted as DP-UniAlCu in the following sections.
The dataset used to train DP-UniAlCu comprises 147,464 structures in total, both at zero and at finite temperatures ranging from 400 K to 1200 K, and includes Cu structures (bulk, surface, and gas-phase nanoparticle), Al$_2$O$_3$ surfaces, and Al$_2$O$_3$-supported Cu nanoparticles.
The dataset is publicly available in a zenodo repository.{\cite{Xu2025Zenodo}}
The uMLIPs examined by this work include DPA2 (mptrj and oc20),\cite{Zhang2024npjComputMater} DPA3 (omat24),\cite{Zhang2025arXiv} MACE-MP (medium),\cite{Batatia2025JCP} MACE-OMAT (medium),\cite{BarrosoLuque2024arXiv} and MatterSim (v1.0.0-1M and v1.0.0-5M).\cite{Yang2024arXiv_mattersim}

\clearpage
\section{Results}
\begin{figure}[hbtp]
    \includegraphics[width=1.00\textwidth]{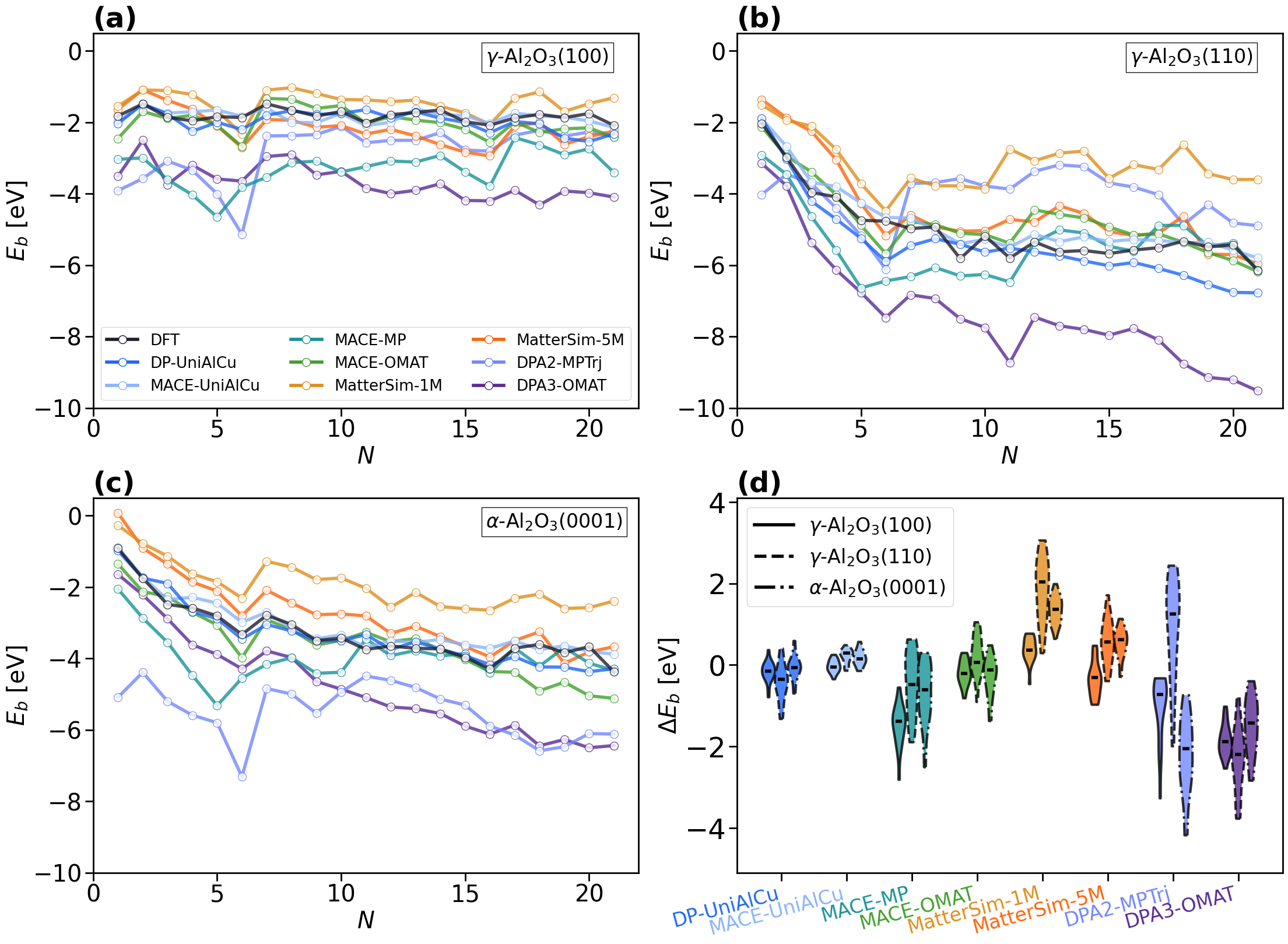}
    \caption{
        Comparison of MLIP-predicted binding energies for Cu$_{1-21}$ supported on three Al$_2$O$_3$ surfaces relative to DFT.
        Binding energies of Cu$_{1-21}$ supported on (a) $\gamma$-Al$_2$O$_3$(100), (b) $\gamma$-Al$_2$O$_3$(110), (c) $\alpha$-Al$_2$O$_3$(0001).
        (d) Violin plots summarizing the distributions of binding-energy errors for Cu$_{1-21}$ predicted by each MLIP relative to DFT.
    }
    \label{fig:binding_energy}
\end{figure}

\textbf{Minimum Energy Configurations from Cu$_\mathbf{1}$ to Cu$_\mathbf{21}$.}
We commence our examination with geometric optimization-based tasks.
The binding energies (defined in \textbf{Equation~\ref{eq:binding_energy}}) of Cu$_{1-21}$ supported on three Al$_2$O$_3$ surfaces are evaluated by various MLIPs.
The corresponding structures were obtained from global optimization in our previous work.\cite{Xu2025ACSNano}
The three Al$_2$O$_3$ surfaces are modeled by $p(3\times2)$ $\gamma$-Al$_2$O$_3$(100), $p(2\times2)$ $\gamma$-Al$_2$O$_3$(110), $p(3\times 2\sqrt{3})$ $\alpha$-Al$_2$O$_3$(0001) periodic slabs (\textbf{Figure~S1}).
The comparison of the absolute total energy is not accessible as these MLIPs are trained on datasets labelled by different DFT calculation settings.
The Perdew-Burke-Ernzerhof (PBE) exchange-correlation functional is used for the reported DFT binding energy results and also the dataset to train our DP-UniAlCu model. 
MACE and MatterSim-branched uMLIPs are trained on datasets labelled by DFT-PBE using a similar setting as ours (\textbf{Table~S1 and S2}), which is the default one named as ``MPRelaxSet'' used in Materials Project.\cite{Jain2013APLMater,Ong2013ComputMaterSci}
The binding energy results from our DFT setting and ``MPRelaxSet'' is almost identical with a maximum absolute error of 0.10 eV (\textbf{Figure~S2}).
Since all MLIPs tested here are trained on DFT-PBE, we are able to examine the accuracy of these MLIPs on the binding energy, where a reliable MLIP should be able to reproduce the binding energy trend across different nanoparticle sizes and surfaces.
We noticed the dispersion interaction may affect the absolute values of binding energies of supported nanoparticles when compared to experiments,\cite{Wellendorff2015SurfSci,Zhao2024ACSCatal} however, the DFT-PBE results presented here only served as a reference for the comparison of different MLIPs and the relative binding strength across different nanoparticle sizes and supporting surfaces is preserved.\cite{Xu2025ACSNano}
\textbf{Figure~\ref{fig:binding_energy}a-c} show the binding energy results of Cu$_{1-21}$ on $\gamma$-Al$_2$O$_3$(100), $\gamma$-Al$_2$O$_3$(110), and $\alpha$-Al$_2$O$_3$(0001), respectively.
The MLIP binding energy results are calculated based on the MLIP total energies of the configurations minimized by the corresponding MLIP model starting from the DFT-minimized configurations.
No significant structural difference between DFT and MLIP-minimized configurations is observed.
Most models give minimized structures with an averaged atomic displacement smaller than $\sim$0.05 \r{A} compared to the DFT-minimized structures except for DPA2-MPTrj and DPA2-OC2M that give displacements larger than $\sim$0.1 \r{A} (\textbf{Figure~S3}).
All examined uMLIPs can reproduce the binding energy trend as a function of the nanoparticle size on a single Al$_2$O$_3$ surface, namely, the binding energy on $\gamma$-Al$_2$O$_3$(100) is almost independent of the nanoparticle size while the binding energies on $\gamma$-Al$_2$O$_3$(110) and $\alpha$-Al$_2$O$_3$(0001) decrease significantly with the nanoparticle size.
The size dependence of the binding energy of the nanoparticles originates from the distinct surface sites present on the three Al$_2$O$_3$ surfaces.
In particular, the presence of strong Al$^{III}$ binding sites on $\gamma$-Al$_2$O$_3$(110) and $\alpha$-Al$_2$O$_3$(0001) (\textbf{Figure~S1}) enables an increasing number of Cu atoms to interact strongly with the support as the nanoparticle size grows, resulting in progressively stronger binding, as discussed in our previous work.{\cite{Xu2025ACSNano}}
According to DFT-PBE, the binding strength follows the trend $\gamma$-Al$_2$O$_3$(110) $>$ $\alpha$-Al$_2$O$_3$(0001) $>$ $\gamma$-Al$_2$O$_3$(100).
The three panels of \textbf{Figure~S4} show the difference in the binding energy of the $N$-atom nanoparticles between pairs of surfaces.
All uMLIPs reproduce the sign of the binding energy difference, with the exception of DPA2-MPTrj and DPA2-OC2M.
For example, DPA2-MPTrj predicts stronger binding on $\alpha$-Al$_2$O$_3$(0001) than on $\gamma$-Al$_2$O$_3$(110), contrary to the DFT-PBE reference.
\textbf{Figure~\ref{fig:binding_energy}d} shows the distribution of binding energy errors through the violin plots, where a good model should have an error distribution centered at zero with a small spread.
The binding energy error is defined as the difference between the binding energy predicted by an MLIP and the binding energy by DFT-PBE.
MACE-OMAT and MatterSim-v1.0.0-5M exhibit small errors with narrow spreads centered near zero, which reproduce both the relationship between the binding energy and the nanoparticle size and the relative binding strength across three Al$_2$O$_3$ surfaces.
It is surprising that MACE-OMAT and MatterSim-v1.0.0-5M are not trained explicitly on any structures of supported nanoparticles but have an accuracy comparable to our DP-UniAlCu model.
MACE-MP, which has the same architecture as MACE-OMAT but is trained on a smaller dataset, is less accurate than MACE-OMAT and predicts more negative binding energies on all three Al$_2$O$_3$ surfaces, which can be seen from the average binding energy errors.
MatterSim-v1.0.0-1M, which is trained on the same dataset as MatterSim-v1.0.0-5M but has a smaller number of model parameters, leads to a degraded accuracy on binding energy, particularly for $\gamma$-Al$_2$O$_3$(110) with the averaged binding energy error of $\sim$2 eV.
The above results indicate that the model performance can be improved either by increasing the dataset size (MACE-MP vs. MACE-OMAT) or by increasing the model size (MatterSim-v1.0.0-1M vs. MatterSim-v1.0.0-5M).
To examine this model improvement strategy, we trained a MACE model based on our UniAlCu dataset using the same architecture as MACE-OMAT, which is denoted as the MACE-UniAlCu.
The details of the training settings are given in \textbf{Section~S3} and the model errors on the entire UniAlCu dataset are shown in \textbf{Figure~S5}.
MACE-UniAlCu exhibits smaller error spreads for all three Al$_2$O$_3$ surfaces compared to both DP-UniAlCu and MACE-OMAT.
The DPA2-MPTrj shows much larger errors compared to other MLIP models.
The large errors are also observed for the DPA2-OC2M (\textbf{Figure~S6}), which is trained on the OC2M dataset.\cite{Chanussot2021ACSCatal}
The DPA3-OMAT exhibits smaller errors than DPA2-MPTrj and DPA2-OC2M due to its improved architecture.
The model architecture can affect the accuracy, as seen from that DPA2-MPTrj is the head trained on the same dataset as MACE-MP, but has larger errors, which is similar in the case of DPA3-OMAT and MACE-OMAT.
In addition, the multi-head training strategy used in DPA2 and DPA3, despite utilizing available datasets to the maximum extent, compromises the accuracy of a specific head due to inconsistent DFT settings in the datasets.

\textbf{Global Optimization of Larger Nanoparticles.}
In practice, the low-energy structures of supported nanoparticles are not available and are usually difficult to construct manually.
Global optimization can explore low-energy structures of supported nanoparticles, where the initial structures are randomly generated, and the final structures can have different geometries but similar energies.
A good MLIP should be able to minimize the initial random structures that may have unphysically short interatomic distances and distinguish the low-energy structures from the high-energy ones.
Therefore, global optimization is a good test for the uMLIPs to examine their stability in minimization and their performance in finding low-energy structures.
Based on the performance of uMLIPs on the binding energies of Cu$_{1-21}$, we examined MACE-OMAT and MatterSim-v1.0.0-1M due to their balance in accuracy and efficiency.
At the same time, we tested DP-UniAlCu as a baseline.
The global optimization based on the genetic algorithm was performed for several larger nanoparticles, including Cu$_{27}$, Cu$_{38}$, Cu$_{47}$ and Cu$_{55}$.
Our UniAlCu training dataset only included supported nanoparticles up to Cu$_{21}$.
Thus, global optimization of larger supported nanoparticles, such as Cu$_{27}$, Cu$_{38}$, Cu$_{47}$, and Cu$_{55}$, and comparison of their energy with that of the corresponding DFT structures is a test of the extrapolation ability of the model.
The periodic slabs of $p(5\times3)$ $\gamma$-Al$_2$O$_3$(100), $p(3\times3)$ $\gamma$-Al$_2$O$_3$(110), $p(5\times3\sqrt3)$ $\alpha$-Al$_2$O$_3$(0001) shown in \textbf{Figure~S7} are used to accommodate the nanoparticles.
In the global optimization task, the initial structures are randomly generated, and the structures in the following 20 generations are generated based on cut-and-splice and mutation algorithms, where the population size is set to 50.
The details of the genetic algorithm are given in \textbf{Section~\nameref{sec:methods}}.
Since the search is finite, the minimum energy structures are unlikely to be the true global minima.
Let $E_b$ be the binding energies of the MLIP minimum energy structures.
After local DFT relaxation the structures differ little from their MLIP counterparts and have energies $E_b^\mathrm{DFT}$. 
For each nanoparticle-surface combination this yields 150 MLIP low energy structures in one to one correspondence with 150 structures locally optimized by DFT.
We define the binding-energy deviation by $\Delta E_b = E_b - E_b^\mathrm{DFT}$.

\textbf{Table~{\ref{tab:comparison_on_global_minima}}} summarizes the energetic metrics of the low-energy structures.
For each model, we report the average binding energy deviation, $\mathrm{avg}(\Delta E_b)$, which should approach zero when the model exactly reproduces DFT, and its standard deviation, $\mathrm{std} (\Delta E_b)$.
These two quantities measure the bias and the spread introduced by the models.
For nanoparticles supported on the two $\gamma$-Al$_2$O$_3$ surfaces, MACE-OMAT yields $\Delta E_b$ values closer to zero among the three models.
These results suggest that MACE-OMAT maintains competitive predictive performance for larger supported nanoparticles, even though it was not explicitly trained on supported nanoparticle structures.
In comparison, MatterSim-v1.0.0-1M has $\Delta E_b$ values comparable to DP-UniAlCu for $\gamma$-Al$_2$O$_3$(100), but makes substantially larger errors for $\gamma$-Al$_2$O$_3$(110).
For $\alpha$-Al$_2$O$_3$(0001), DP-UniAlCu provides the best accuracy, while MACE-OMAT and MatterSim-v1.0.0-1M exhibit larger deviations.
Overall, each model achieves its best performance on a different support: DP-UniAlCu on $\alpha$-Al$_2$O$_3$(0001), MACE-OMAT on $\gamma$-Al$_2$O$_3$(110), and MatterSim-v1.0.0-1M on $\gamma$-Al$_2$O$_3$(100).
This non-uniform accuracy indicates that the extrapolation performance of the MLIPs is not consistent across different supports, likely reflecting differences in training data coverage and protocols.
To further examine this behavior, we locally relaxed the above 150 structures with MACE-UniAlCu to obtain the corresponding $\Delta E_b$. 
The energetic metrics obtained in this way are reported in \textbf{Table~S3}, which shows that MACE-UniAlCu exhibits improved accuracy for $\gamma$-Al$_2$O$_3$(100) but reduced accuracy for the other two surfaces relative to DP-UniAlCu, reinforcing the conclusion that extrapolation performance varies across systems even for models trained on the same dataset.

In practice, MLIP-generated low-energy structures can be further minimized by DFT to identify the DFT low-energy structures.
Therefore, the ability of an MLIP not only to include but also to preserve the relative energetic ordering of these DFT low-energy structures serves as a key indicator of its performance.
Two metrics (\textbf{Table~\ref{tab:ranking_metrics}}) are used to quantify the quality of MLIP-generated low-energy structures, namely, rank biased overlap\cite{Webber2010ACMTOIS} (RBO, defined in \textbf{Equation~\ref{eq:rbo}}) and Kendall's $\tau$\cite{Kendall1945Biometrika} (defined in \textbf{Equation~\ref{eq:kendalltau}}).
The RBO measures the similarity between two lists of binding energies, the DFT ones and the MLIP ones, and focuses more on the top of the lists, where a larger RBO value indicates a better MLIP performance.
However, it is less sensitive to the order of the binding energies in the lists.
In addition, Kendall's $\tau$ measures the ordinal association between two lists of binding energies, where a larger value indicates a better MLIP performance.
In seven out of twelve systems, our DP-UniAlCu model shows the best performance in both RBO and Kendall's $\tau$, indicating its excellent ability to generate low-energy structures consistent with DFT.
For the rest of the systems, MACE-OMAT performs better than DP-UniAlCu.
MatterSim-v1.0.0-1M gives less satisfactory results and only exhibits comparable performance on a few systems, such as Cu$_{47}$ on $\gamma$-Al$_2$O$_3$(100) and Cu$_{55}$ on $\gamma$-Al$_2$O$_3$(110).

\begin{figure}[!hbtp]
    \includegraphics[width=1.00\textwidth]{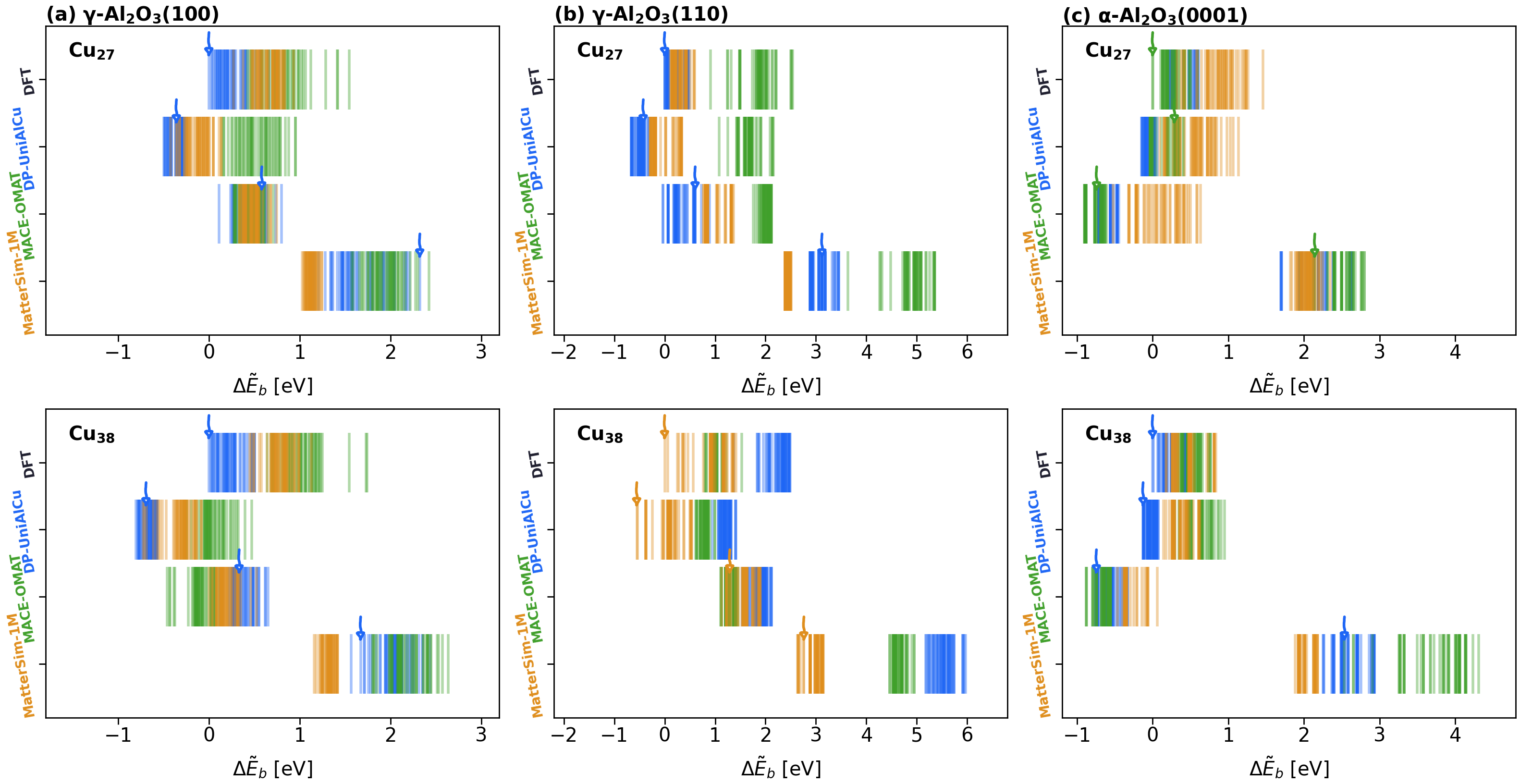}
    \caption{
        Comparison of MLIP-predicted and DFT-refined binding-energy spectra for supported Cu$_{27}$ and Cu$_{38}$ nanoparticles on Al$_2$O$_3$ surfaces.
        Low-energy structures are ranked by the deviation of their binding energies ($E_b$) relative to the minimum binding energy among DFT-minimized structures ($\min(E_b^\mathrm{DFT})$), defined as $\Delta \tilde E_b = E_b - \min(E_b^{\mathrm{DFT}})$.
        The six panels correspond to Cu$_{27}$ and Cu$_{38}$ nanoparticles supported on three Al$_2$O$_3$ surfaces: Cu$_{27}$/$\gamma$-Al$_2$O$_3$(100) and Cu$_{38}$/$\gamma$-Al$_2$O$_3$(100) in (a); Cu$_{27}$/$\gamma$-Al$_2$O$_3$(110) and Cu$_{38}$/$\gamma$-Al$_2$O$_3$(110) in (b); and Cu$_{27}$/$\alpha$-Al$_2$O$_3$(0001) and Cu$_{38}$/$\alpha$-Al$_2$O$_3$(0001) in (c).
        Each panel contains four spectra. The three lower spectra show $\Delta E_b$ predicted by DP-UniAlCu, MACE-OMAT, and MatterSim-v1.0.0-1M, respectively.
        Vertical bars are color-coded according to the model used to generate the structures: DP-UniAlCu (blue), MACE-OMAT (green), and MatterSim-v1.0.0-1M (orange).
        The top spectrum reports $\Delta E_b$ obtained after local DFT minimization of structures generated by the MLIP models (color-coded as above).
        Vertical arrows in the top spectrum indicate the minimum DFT-minimized binding energy, $\min(E_b^{\mathrm{DFT}})$, for each system.
        The corresponding $\Delta \tilde E_b$ predicted by DP-UniAlCu, MACE-OMAT, and MatterSim-v1.0.0-1M for that structure are marked by arrows in the three lower spectra.
    }
    \label{fig:minimum_spectrum_00}
\end{figure}

\begin{figure}[!hbtp]
    \includegraphics[width=1.00\textwidth]{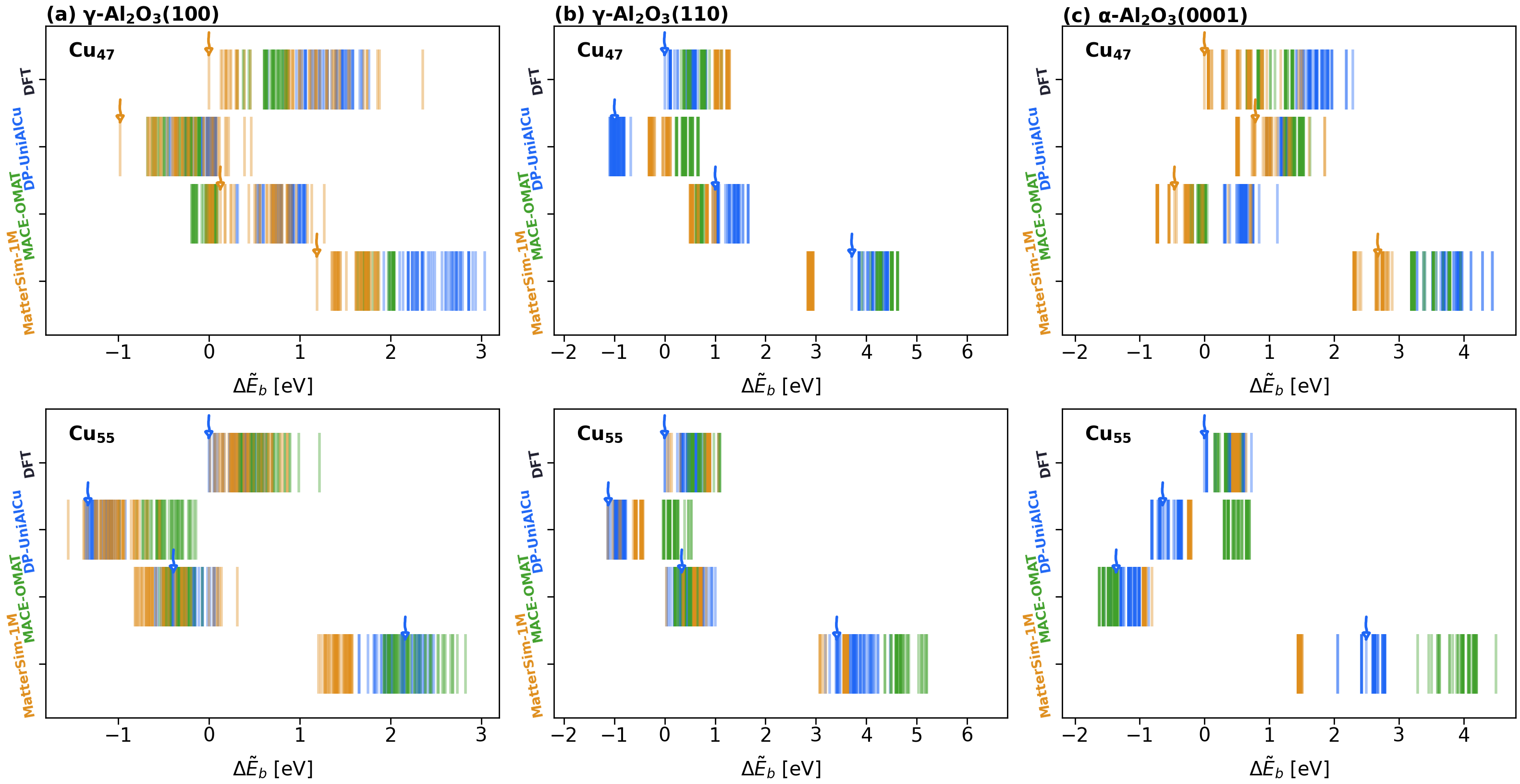}
    \caption{
        Comparison of MLIP-predicted and DFT-refined binding-energy spectra for supported Cu$_{47}$ and Cu$_{55}$ nanoparticles on Al$_2$O$_3$ surfaces.
        Low-energy structures are ranked by the deviation of their binding energies ($E_b$) relative to the minimum binding energy among DFT-minimized structures ($\min(E_b^\mathrm{DFT})$), defined as $\Delta \tilde E_b = E_b - \min(E_b^{\mathrm{DFT}})$.
        The six panels correspond to Cu$_{47}$ and Cu$_{55}$ nanoparticles supported on three Al$_2$O$_3$ surfaces: Cu$_{47}$/$\gamma$-Al$_2$O$_3$(100) and Cu$_{55}$/$\gamma$-Al$_2$O$_3$(100) in (a); Cu$_{47}$/$\gamma$-Al$_2$O$_3$(110) and Cu$_{55}$/$\gamma$-Al$_2$O$_3$(110) in (b); and Cu$_{47}$/$\alpha$-Al$_2$O$_3$(0001) and Cu$_{55}$/$\alpha$-Al$_2$O$_3$(0001) in (c).
        Each panel contains four spectra. The three lower spectra show $\Delta E_b$ predicted by DP-UniAlCu, MACE-OMAT, and MatterSim-v1.0.0-1M, respectively.
        Vertical bars are color-coded according to the model used to generate the structures: DP-UniAlCu (blue), MACE-OMAT (green), and MatterSim-v1.0.0-1M (orange).
        The top spectrum reports $\Delta E_b$ obtained after local DFT minimization of structures generated by the MLIP models (color-coded as above).
        Vertical arrows in the top spectrum indicate the minimum DFT-minimized binding energy, $\min(E_b^{\mathrm{DFT}})$, for each system.
        The corresponding $\Delta \tilde E_b$ predicted by DP-UniAlCu, MACE-OMAT, and MatterSim-v1.0.0-1M for that structure are marked by arrows in the three lower spectra.
    }
    \label{fig:minimum_spectrum_01}
\end{figure}

To better visualize the ability of the MLIP models to identify low-energy structures across nanoparticle sizes, for each system (nanoparticle-surface) we rank all candidate structures by their binding-energy deviation relative to the minimum energy structure among the 150 DFT-optimized structures, i.e., $\Delta \tilde E_b = E_b - \min(E_b^\mathrm{DFT})$. 
The $\Delta \tilde E_b$ spectra of the nanoparticles on the three Al$_2$O$_3$ surfaces are shown in \textbf{Figure~{\ref{fig:minimum_spectrum_00}}a-c} for Cu$_{27}$ and Cu$_{38}$ and in \textbf{Figure~{\ref{fig:minimum_spectrum_01}}a-c} for Cu$_{47}$ and Cu$_{55}$.
For each system, the three spectra at the bottom of a panel show the $\Delta \tilde E_b$ values predicted by DP-UniAlCu, MACE-OMAT, and MatterSim-v1.0.0-1M, respectively, while the spectrum at the top of the panel reports the corresponding binding energies after local DFT optimization.
The structures are color-coded according to the MLIP used for their generation, that is blue for DP-UniAlCu, green for MACE-OMAT, and orange for MatterSim-v1.0.0-1M.
In each panel the minimum DFT binding energy ( $\min(E_b^\mathrm{DFT})$) is aligned with the zero of the energy scale and is indicated by a vertical arrow.
The arrows in the MLIP spectra indicate the MLIP energy of the structure corresponding to the minimum DFT structure.

DP-UniAlCu finds the structure of minimum $E_b$ in eight cases, MatterSim-v1.0.0-1M in three cases, and MACE-OMAT in one case.
While the DP-UniAlCu and MACE-OMAT spectra overlap significantly with each other and with the DFT-minimized spectrum, MatterSim-v1.0.0-1M is systematically shifted to higher $E_b$ values.
Further analysis of the binding-energy error decomposition (\textbf{Table~S4}) shows that the dominant binding-energy errors in MatterSim-v1.0.0-1M originate from the gas-phase nanoparticle energies.
Since optimization of supported nanoparticles is performed at fixed composition, these errors introduce a constant offset and therefore do not affect the relative energy ranking among low-energy structures.
In the four cases in which DP-UniAlCu fails to locate the structure with the lowest DFT energy, the corresponding $E_b$ is very close to $\min E_b^\mathrm{DFT}$.
Interestingly, the DP energy of two minimum energy structures identified by MatterSim-v1.0.0-1M is also the minimum energy of the DP spectrum, indicating that the correct identification of low energy structures depends not only on the accuracy of the MLIP but also reflects the finite character and stochastic nature of a single search by the genetic algorithm.
Although MatterSim-v1.0.0-1M exhibits a systematic binding energy shift, the minimum energy structures that it identifies have good ranking metrics (RBO and Kendall's $\tau$), similar to those of DP-UniAlCu, indicating that these structures are not a random outcome of a search.
This suggests that there is no strong correlation between energetic predictions and ranking metrics.
To further test this hypothesis we report in \textbf{Table~S5} the ranking metrics of MACE-UniAlCu on the same low-energy structures. While the energy accuracy of MACE-UniAlCu is sometimes better and sometimes worse than that of DP-UniAlCu (see \textbf{Table~S3}), MACE-UniAlCu achieves consistently improved ranking metrics for almost all the systems.
These results suggest that even in the presence of significant absolute binding-energy errors a model can still predict a meaningful ordering of the low energy structures.
To improve our confidence on the predicted structures it is useful to complement energetic predictions with ranking metrics.

\textbf{Finite-Temperature Dynamics.}
The dynamic behavior of supported nanoparticles is important for understanding the catalytic activity and stability.
MD trajectories are driven by forces, and we report in \textbf{Table~{\ref{tab:model_performance_on_unialcu}}} the root mean-squared errors (RMSEs) of the MLIP forces relative to their DFT counterpart, using the UniAlCu dataset, which includes three subsystems, i.e., pure Cu structures, $\gamma$-Al$_2$O$_3$ surfaces with and without supported nanoparticles, and $\alpha$-Al$_2$O$_3$ surfaces with and without supported nanoparticles.
Unsurprisingly, DP-UniAlCu and MACE-UniAlCu give the best force predictions.
MACE-OMAT, DPA3-OMAT and MatterSim-v1.0.0-1M show similar accuracy.
Several MD-based tasks are designed to examine MLIPs, as discussed below.

\begin{figure}[hbtp]
    \includegraphics[width=1.00\textwidth]{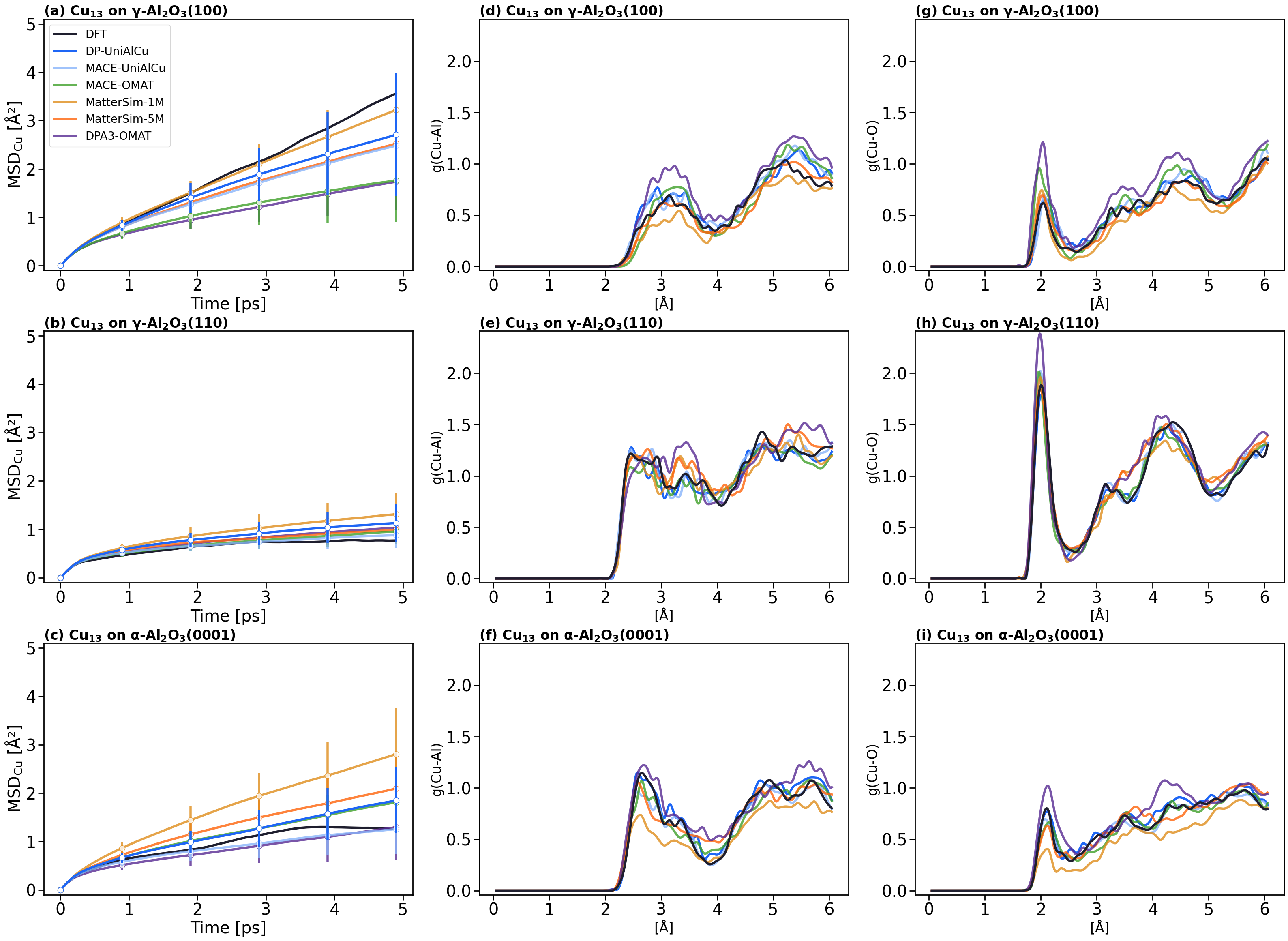}
    \caption{
        Molecular dynamics validation of MLIPs for Cu$_{13}$ supported on Al$_2$O$_3$ at 800 K.
        For each MLIP, 20 MD simulations starting from the same minimum energy configuration of Cu$_{13}$ but initialized from different velocities by Maxwell-Boltzmann distribution at 800 K are performed for 20 ps with a timestep of 2 fs.
        For AIMD, only one trajectory is performed with the same settings.
        (a)-(c) The mean square displacement of Cu atoms.
        The vertical bars are the standard deviations of 20 MD simulations by MLIPs.
        (d)-(f) The radial distribution functions of Cu-Al.
        (g)-(i) The radial distribution functions of Cu-O.
    }
    \label{fig:short_dynamics}
\end{figure}

We first tested the mean-squared displacements of Cu atoms (MSD$_\mathrm{Cu}$, defined in \textbf{Equation~\ref{eq:msd}}) of Cu$_{13}$ on three Al$_2$O$_3$ surfaces with 20-ps MD simulations, where \textit{ab initio} MD (AIMD) trajectories are available for reference.
The DP-UniAlCu, MACE-UniAlCu, MACE-OMAT, MatterSim-v1.0.0-1M, MatterSim-v1.0.0-5M, and DPA3-OMAT are tested.
For each MLIP, 20 MD simulations starting from the same minimum energy configuration but initialized from different velocities by Maxwell-Boltzmann distribution at 800 K are performed for 20 ps with a timestep of 2 fs.
For AIMD, only one trajectory is performed with the same settings.
The window-averaged MSD$_\mathrm{Cu}$ up to 5 ps are shown in \textbf{Figure~\ref{fig:short_dynamics}a-c}.
The vertical bars are the standard deviations of 20 MD simulations by MLIPs.
The model accuracy can be examined by checking whether the AIMD results fall within the standard deviations of MLIPs.
For $\gamma$-Al$_2$O$_3$(100), all models except for MACE-OMAT and DPA3-OMAT give MSD$_\mathrm{Cu}$ results that are close to the AIMD results, which exhibit smaller MSD$_\mathrm{Cu}$ values.
For $\gamma$-Al$_2$O$_3$(110), all models give consistent results with AIMD.
For $\alpha$-Al$_2$O$_3$(0001), all models except for two MatterSim models produce MSD$_\mathrm{Cu}$ results closer to AIMD, which show larger MSD$_\mathrm{Cu}$ values.
The radial distribution functions (RDFs) of Cu-Al and Cu-O pairs are shown in \textbf{Figure~\ref{fig:short_dynamics}d-f} and \textbf{Figure~\ref{fig:short_dynamics}g-i}, respectively, which are averaged over snapshots every 0.1 ps from 0.2 ps to 20 ps.
The RDFs of Cu-Al and Cu-O by DP-UniAlCu, MACE-UniAlCu and MACE-OMAT are in good agreement with the AIMD results, while other models give larger deviations.
The RDF of Cu-Cu can be well reproduced by all models (\textbf{Figure~S8}). 
In addition to the above models, we also tested DPA2-MPTrj and DPA2-OC2M.
However, MD simulations with these models were not stable and collapsed within a few ps at 800 K, consistent with the large force errors on the UniAlCu dataset shown in \textbf{Table~S6}.

\begin{figure}[p]
    \includegraphics[width=1.00\textwidth]{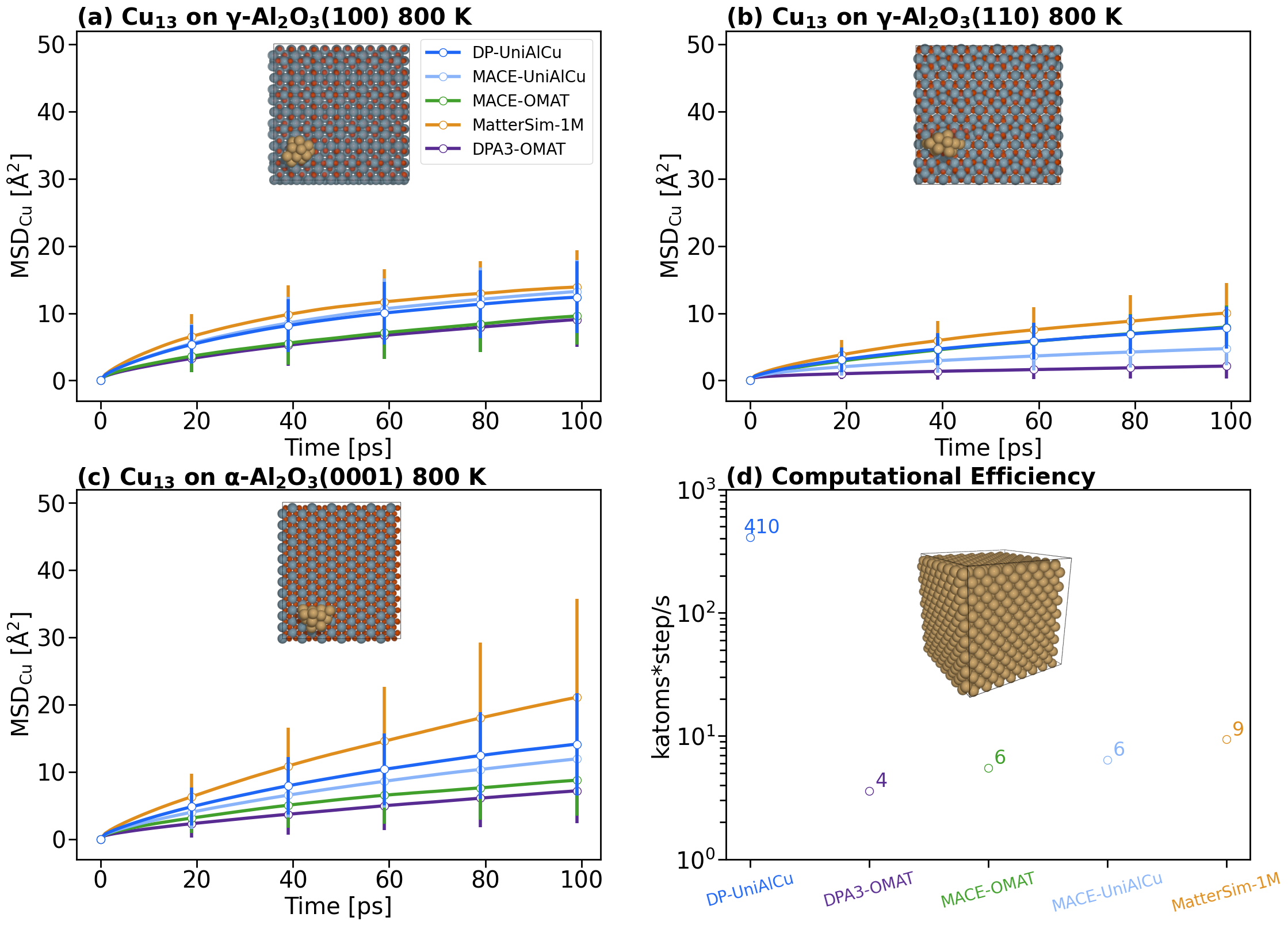}
    \caption{
        Mean square displacement and computational performance of MLIPs.
        (a)-(c) The mean square displacement of Cu atoms for Cu$_{13}$ on three Al$_2$O$_3$ surfaces at 800 K.
        (d) The computational efficiency of DP-UniAlCu, DPA3-OMAT, MACE-OMAT, MACE-UniAlCu, and MatterSim-v1.0.0-1M models on the simulation of an FCC Cu bulk with 2048 atoms.
        The ``katoms*step/s'' is defined as the number of thousands of atoms processed per MD step per second. 
        The inset panels show the initial structures of MD simulations.
    }
    \label{fig:long_dynamics}
\end{figure}

Since the 20-ps-long MD simulations are too short to observe any diffusion of the entire nanoparticle, we performed 1-ns-long MD simulations for Cu$_{13}$ on three $\alpha$-Al$_2$O$_3$(0001) surfaces with DP-UniAlCu, MACE-UniAlCu, MACE-OMAT, MatterSim-v1.0.0-1M, and DPA3-OMAT to investigate the long-time dynamic behavior.
Five independent MD simulations are performed for each MLIP starting from the same minimum energy configuration but initialized from different velocities by Maxwell-Boltzmann distribution at 800 K.
The results of MSD$_\mathrm{Cu}$ up to 100 ps for Cu$_{13}$ on three Al$_2$O$_3$ surfaces are shown in \textbf{Figure~\ref{fig:long_dynamics}a-c}.
The vertical bars are the standard deviations of five MD simulations by MLIPs.
Since ns-long MD simulations are not readily accessible to AIMD, the benchmark here only served as a cross-validation across different MLIPs by comparing results against our DP-UniAlCu model.
All the above MLIPs predict qualitatively consistent mobility trends for the Cu atoms in the nanoparticles supported on the three Al$_2$O$_3$ surfaces.
The mobility of Cu$_{13}$ on these three surfaces was discussed in previous work,{\cite{Xu2025ACSNano}} where it was found that MSD$_\mathrm{Cu}$ on $\gamma$-Al$_2$O$_3$(100) and $\alpha$-Al$_2$O$_3$(0001) are comparable and larger than on $\gamma$-Al$_2$O$_3$(110).
MatterSim-v1.0.0-1M gives slightly larger MSD$_\mathrm{Cu}$ for Cu$_{13}$ on $\gamma$-Al$_2$O$_3$(100) and $\gamma$-Al$_2$O$_3$(110), and much larger MSD$_\mathrm{Cu}$ for $\alpha$-Al$_2$O$_3$(0001) compared to DP-UniAlCu.
MACE-OMAT gives almost identical MSD$_\mathrm{Cu}$ for Cu$_{13}$ on $\gamma$-Al$_2$O$_3$(110) as DP-UniAlCu while more deviations between them are observed for Cu$_{13}$ on $\gamma$-Al$_2$O$_3$(100) and $\alpha$-Al$_2$O$_3$(0001), where MACE-OMAT gives smaller MSD$_\mathrm{Cu}$ values.
By using the MACE-UniAlCu model, which is trained on the same dataset as DP-UniAlCu, the MSD$_\mathrm{Cu}$ results for $\gamma$-Al$_2$O$_3$(100) and $\alpha$-Al$_2$O$_2$(0001) become closer to those by DP-UniAlCu, however, a smaller MSD$_\mathrm{Cu}$ is observed for $\gamma$-Al$_2$O$_3$(110).
DPA3-OMAT exhibits a consistent mobility trend as DP-UniAlCu but gives smaller MSD$_\mathrm{Cu}$ values for all three Al$_2$O$_3$ surfaces.
To perform a more quantitative comparison, more simulations with a longer time are necessary, which can be challenging for uMLIPs.
The computational efficiency is shown in \textbf{Figure~\ref{fig:long_dynamics}d}, which is taken from the 100-ps simulation of an FCC Cu bulk with 2048 atoms.
All models are performed on a single NVIDIA 80G A100 GPU with LAMMPS\cite{Thompson2022ComputPhysCommun} except for MatterSim-v1.0.0-1M with ASE\cite{Larsen2017JPhysCondensMatter}.
DP-UniAlCu is roughly two orders of magnitude faster than the DPA3-OMAT and MACE-based models and more than one order of magnitude faster than MatterSim-v1.0.0-1M.
The uMLIPs and domain-specific MLIPs serve different purposes.
Domain-specific applications that require large size simulations in space and/or time may be only accessible with domain-specific models with available computational resources.

\clearpage
\section{Discussion}
In this work, we tested several uMLIPs on supported Cu nanoparticles on Al$_2$O$_3$ surfaces and found that they can be very useful in simulating supported nanoparticles even without any fine-tuning, including the global optimization of low-energy structures and the MD at finite temperature.

The uMLIPs examined in this work can produce reasonable results on the binding energies of small supported nanoparticles (Cu$_\mathrm{1-21}$) reported in our previous study,\cite{Xu2025ACSNano} where MACE-OMAT shows an accuracy comparable to our domain-specific DP-UniAlCu model.
In the global optimization of large supported nanoparticles (Cu$_\mathrm{27}$, Cu$_\mathrm{38}$, Cu$_\mathrm{47}$, and Cu$_\mathrm{55}$), our domain-specific DP-UniAlCu model more frequently locates low-energy structures than MACE-OMAT and Mattersim-v1.0.0-1M.
Interestingly, Mattersim-v1.0.0-1M, which shows larger deviations in binding energies, can find even more stable configurations than the other two models in some systems, indicating its exploration ability in the configuration space is not solely characterized by the energetic metrics on a preset dataset.
Alongside the exploration ability of MatterSim-v1.0.0-1M, the non-uniform accuracy across different systems is observed, for instance, the binding energy accuracy in one nanoparticle size is better than another size on the same Al$_2$O$_3$ surface.
This non-uniform accuracy may be due to the imbalance in the number of configurations from different systems, where the initial dataset is made up of configurations from one specific system and the configurations of other systems are accumulated in active learning.
In terms of global optimization, since the system size of interest is usually accessible to DFT, the MLIP-generated configurations can still be verified by DFT to find the 
low-energy structures.
The model's stability in generating diverse configurations in a large region of the configuration space is more important than the accuracy of the energy, as it provides more candidates for DFT verification.
A model with better accuracy can improve the performance in the workflow of global optimization since it can produce more low-energy structures close to DFT and reduce the number of DFT calculations, however, the improved accuracy generally comes with an increased computational cost. 
Though minimization is fast for a single structure and can be made parallel, the computational cost can become intensive for complex systems even by MLIPs, where numerous structures are calculated.

The MD benchmark for supported nanoparticles shows that uMLIPs can produce results consistent with AIMD for MSD$_\mathrm{Cu}$ and RDF in simulations having a 20 ps timescale.
When larger scale MD simulations are required, benchmarking becomes challenging in two aspects: the lack of DFT reference and the computational cost.
The properties predicted by the MLIPs in the long-time limit can only be cross-validated within the models in the absence of AIMD.
MACE-OMAT and MatterSim-v1.0.0-1M exhibit qualitative trends for MSD$_\mathrm{Cu}$ up to 100 ps similar to the DP-UniAlCu and MACE-UniAlCu models, but this is still too short to reliably estimate
MSD$_\mathrm{Cu}$. Longer MD simulations are necessary for that, but they may not be feasible with uMLIPs.
DP-UniAlCu is about two orders of magnitude faster than MACE-OMAT and Mattersim-v1.0.0-1M. The latter models are not only universal, but also have better accuracy for energy and forces due to their more advanced architectures.
However, the energy and force accuracy of the standard DP model can be improved with marginal additional cost by adopting message passing and a slightly higher level of equivariance ~\cite{Gao2024PCCP}.
uMLIPs can be very useful in the development of system specific MLIPs particularly in the initial stages of construction of the training dataset, when this is typically not sufficiently structurally diverse.{\cite{Zhang2019PRM}}
For this purpose, DFT labeling can be replaced by uMLIP labeling, gaining both structural diversity and computational cost. DFT labeling can then be utilized more rarely only in the final stages of the construction of the data set with a substantial gain in computational cost. 
Ideally, a uMLIP could be distilled into an efficient model specific MLIP without any DFT labeling.
Our work shows that this procedure can be problematic because the accuracy of the uMLIPs may be not uniform, and they can occasionally generate unphysical structures.   
In this context, a possible way to be explored would require using uncertainty quantification techniques, such as discussed in Refs. {\citenum{Bilbrey2025npjComputMater,Liu2025npjComputMater}}, as a diagnostic step to assess whether a given uMLIP reliably describes the relevant configuration space before distillation.
The benchmarks discussed in this work should provide useful insight on how uMLIPs could be used in application studies of supported nanoparticle catalysts.

\section{Methods}\label{sec:methods}
\subsection{Density Functional Theory and Machine Learning Interatomic Potentials}
The DFT calculations used in binding energy computation and ab initio molecular dynamics are performed with the Vienna Ab initio Simulation Package (VASP).\cite{Kresse1996ComputMaterSci,Kresse1996PRB}
The Perdew-Burke-Ernzerhof (PBE) approximation for the exchange-correlation functional\cite{Perdew1996PRL} is used in our UniAlCu dataset and in calculating the binding energies of supported nanoparticles.
The MACE-MP, MACE-OMAT, MatterSim-v1.0.0-1M, and MatterSim-v1.0.0-5M are trained on the PBE-based datasets with slightly different parameters.
The detailed comparison of input settings can be found in \textbf{Section~S2}.
The MACE-UniAlCu model is trained on our UniAlCu dataset using the same setting as MACE-OMAT, which is discussed in \textbf{Section~S3}.

\subsection{Molecular Dynamics and Global Optimization}
The simulations by uMLIPs are performed with either LAMMPS\cite{Thompson2022ComputPhysCommun} or ASE\cite{Larsen2017JPhysCondensMatter} through interfaces implemented in the GDPy package.\cite{Xu2024GDPy}
The global optimization is based on the genetic algorithm implemented in the GDPy package,\cite{Lee2022JPCC,Xu2024GDPy} where the crossover used a cut-and-splice algorithm\cite{Deaven1995PRL,Vilhelmsen2012PRL} and the mutations include the rattle and mirror operations with equal probabilities.
The population size is set to 50 and the number of generations is set to 20.
The configurations in the first generation are randomly generated, and those in the following generations are produced by the crossover and mutation methods.

\subsection{Benchmarking Metrics}
The binding energy $E_b$ of a supported nanoparticle is defined as
\begin{equation}\label{eq:binding_energy}
    E_b = E_{{\mathrm{Cu}_N}/surf} - E_{surf} - E_{{\mathrm{Cu}_N}}
\end{equation}
where $E_{\mathrm{Cu}_N/surf}$ is the potential energy of a Cu nanoparticle on an Al$_2$O$_3$ surface, $E_{surf}$ is the potential energy of a pristine Al$_2$O$_3$ surface, and $E_{\mathrm{Cu}_N}$ is the potential energy of the minimum energy configuration of Cu$_N$ in the gas phase.

Two ranking metrics are used to quantify the quality of MLIP-generated low-energy structures.
The RBO\cite{Webber2010ACMTOIS} is defined as
\begin{equation}\label{eq:rbo}
    q(S,T,p) = (1-p)\sum_{d=1}^{\infty} p^{d-1} \frac{\lvert S_{1:d} \cap T_{1:d} \rvert}{d},
\end{equation}
where $S$ and $T$ are two lists of binding energies ranked from low to high, $d$ is the length of a sub-list, and $p$ is a parameter controlling the top-weightness and chosen to be 0.94 when 95\% of the weight is contributed by the top fifty energies.
The Kendall's $\tau$\cite{Kendall1945Biometrika} is defined as
\begin{equation}\label{eq:kendalltau}
    \tau = \frac {P - Q}{\sqrt{(P + Q + T)(P + Q + U)}}
\end{equation}
where $P$ is the number of concordant pairs, $Q$ is the number of discordant pairs, $T$ is the number of ties only in the first list, and $U$ is the number of ties only in the second list.

The MSD$_\mathrm{Cu}$ is given by
\begin{equation}\label{eq:msd}
    \mathrm{MSD}(t;r) = {<\frac{1}{N}\sum_{i=1}^{N} \lVert r(t+\tau)-r(\tau)\rVert^2>}_{\tau},
\end{equation}
where $N$ is the number of Cu atoms, $r(t)$ is the coordinate at timestep $t$, $\tau$ is the time origin, and $<\dotsb>_{\tau}$ denotes the average over all time origins with a lagtime in one trajectory.

\clearpage
\section{Data Availability}
The UniAlCu dataset is available online,{\cite{Xu2025Zenodo}} with information given in the respective publication.\cite{Xu2025ACSNano}

\section{Acknowledgements}
The funding for this project was provided by Shell International Exploration and Production Inc., USA.
The calculations were performed largely using the Princeton Research Computing resources at Princeton University.
The authors also acknowledge the computing resources from the National Energy Research Scientific Computing Center (NERSC) operated under Contract No. DE-AC0205CH11231 using NERSC award ERCAP0021510.
The software used in the present work has been developed in the Computational Chemical Science Center “Chemistry in Solution and at Interfaces (CSI)” funded by the USA Department of Energy under award DE-SC0019394.
R.C. was partially supported by this grant.

\section{Supplementary Information}
8 figures, 6 tables, notes and references

\section{Author contributions}
J.X., A.P., A.D.P., S.S., and R.C. conceived the idea of this work and designed the benchmark.
J.X. performed the calculations and wrote the original draft.
J.X., A.P., S.S., and R.C. revised the manuscript.
J.X., A.P., A.D.P., S.S., D.H., and R.C. contributed to the discussions and reviewed the manuscript.

\section{Competing interests}
The authors declare no competing financial or non-financial interests.

\clearpage
\bibliography{references}

\clearpage

\begin{table}[hbtp]
    \caption{Energetic Metrics on Low-Energy Structures (units in eV)}
    \label{tab:comparison_on_global_minima}
    \makebox[\textwidth][c]{
    \resizebox{1.2\textwidth}{!}
    {
    \begin{tabular}{l l r r r r r r}
        \hline
        \multicolumn{2}{c}{Model} & 
        \multicolumn{2}{c}{DP-UniAlCu} & 
        \multicolumn{2}{c}{MACE-OMAT} & 
        \multicolumn{2}{c}{MatterSim-v1.0.0-1M} \\
        Surface & Size & 
        $avg(\Delta E_b) \rightarrow 0$ & $std(\Delta E_b)\downarrow$ & 
        $avg(\Delta E_b) \rightarrow 0$ & $std(\Delta E_b)\downarrow$ & 
        $avg(\Delta E_b) \rightarrow 0$ & $std(\Delta E_b)\downarrow$ \\
        \hline
        $\gamma$-Al$_2$O$_3$(100)   & Cu$_{27}$ &         -0.56   &  0.27  &   \textbf{-0.11} &  0.28  &   1.01  &  0.48  \\
                                    & Cu$_{38}$ &         -0.99   &  0.22  &   \textbf{-0.53} &  0.54  &   1.17  &  0.55  \\
                                    & Cu$_{47}$ &         -1.21   &  0.32  &   \textbf{-0.55} &  0.25  &   0.99  &  0.40  \\
                                    & Cu$_{55}$ &         -1.33   &  0.26  &   \textbf{-0.80} &  0.19  &   1.48  &  0.35  \\
        $\gamma$-Al$_2$O$_3$(110)  & Cu$_{27}$ &          -0.42   &  0.30  &   \textbf{0.31}  &  0.38  &   2.68  &  0.42  \\
                                    & Cu$_{38}$ &         -0.70   &  0.39  &   \textbf{0.20}  &  0.45  &   2.99  &  0.72  \\
                                    & Cu$_{47}$ &         -0.96   &  0.55  &   \textbf{0.13}  &  0.55  &   3.08  &  0.93  \\
                                    & Cu$_{55}$ &         -1.11   &  0.45  &   \textbf{-0.15} &  0.16  &   3.34  &  0.55  \\
         $\alpha$-Al$_2$O$_3$(0001) & Cu$_{27}$ & \textbf{-0.27}  &  0.20  &          -0.94   &  0.20  &   1.72  &  0.49  \\
                                    & Cu$_{38}$ & \textbf{-0.11}  &  0.23  &          -0.94   &  0.22  &   2.36  &  0.74  \\
                                    & Cu$_{47}$ & \textbf{ 0.07}  &  0.47  &          -1.05   &  0.23  &   2.17  &  0.31  \\
                                    & Cu$_{55}$ & \textbf{-0.51}  &  0.47  &          -1.61   &  0.19  &   2.25  &  1.08  \\
    \end{tabular}
    }
    }
\end{table}

\clearpage
\begin{table}[hbtp]
    \caption{Ranking Metrics on Low-Energy Structures}
    \label{tab:ranking_metrics}
    \makebox[\textwidth][c]{
    \resizebox{1.2\textwidth}{!}
    {
    \begin{tabular}{l l c r r r}
        \hline
        Surface & 
        Size &
        \begin{tabular}{c c}RBO q$\uparrow$ \\ Kendall $\tau$$\uparrow$\end{tabular}&
        DP-UniAlCu & 
        MACE-OMAT & 
        MatterSim-v1.0.0-1M \\
        \hline
        $\gamma$-Al$_2$O$_3$(100)   &Cu$_{27}$  & q      & \textbf{ 0.36}     &          0.29      &           0.02  \\
                                    &           & $\tau$ & \textbf{ 0.58}     &          0.26      &           0.05  \\
                                    &Cu$_{38}$  & q      & \textbf{ 0.41}     &          0.01      &           0.02  \\
                                    &           & $\tau$ & \textbf{ 0.63}     &         -0.43      &           0.07  \\
                                    &Cu$_{47}$  & q      &          0.70      &          0.21      &  \textbf{ 0.77} \\
                                    &           & $\tau$ &          0.51      & \textbf{ 0.70}     &           0.46  \\
                                    &Cu$_{55}$  & q      & \textbf{ 0.44}     &          0.30      &           0.26  \\
                                    &           & $\tau$ & \textbf{ 0.48}     & \textbf{ 0.48}     &           0.39  \\
        $\gamma$-Al$_2$O$_3$(110)   &Cu$_{27}$  & q      & \textbf{ 0.37}     &          0.13      &           0.06  \\
                                    &           & $\tau$ & \textbf{ 0.60}     &          0.52      &           0.46  \\
                                    &Cu$_{38}$  & q      & \textbf{ 0.75}     &          0.30      &           0.52  \\
                                    &           & $\tau$ & \textbf{ 0.59}     & \textbf{ 0.59}     &           0.56  \\
                                    &Cu$_{47}$  & q      & \textbf{ 0.15}     &          0.02      &           0.01  \\
                                    &           & $\tau$ & \textbf{ 0.21}     &         -0.27      &          -0.15  \\
                                    &Cu$_{55}$  & q      &          0.46      &          0.33      &  \textbf{ 0.49} \\
                                    &           & $\tau$ &          0.26      & \textbf{ 0.56}     &          -0.01  \\
        $\alpha$-Al$_2$O$_3$(0001)  &Cu$_{27}$  & q      &          0.14      & \textbf{ 0.28}     &           0.01  \\
                                    &           & $\tau$ &          0.50      & \textbf{ 0.62}     &          -0.20  \\
                                    &Cu$_{38}$  & q      & \textbf{ 0.54}     &          0.13      &           0.05  \\
                                    &           & $\tau$ & \textbf{ 0.42}     &          0.12      &           0.05  \\
                                    &Cu$_{47}$  & q      &          0.45      & \textbf{ 0.52}     &           0.48  \\
                                    &           & $\tau$ &          0.27      & \textbf{ 0.81}     &           0.61  \\
                                    &Cu$_{55}$  & q      & \textbf{ 0.43}     &          0.25      &           0.01  \\
                                    &           & $\tau$ &         -0.06      & \textbf{ 0.43}     &          -0.21  \\
    \end{tabular}
    }
    }
\end{table}

\clearpage

\clearpage
\begin{table}[hbtp]
    \centering
    \caption{Force Root Mean-Squared Errors on UniAlCu Dataset}
    \label{tab:model_performance_on_unialcu}
    \makebox[\textwidth][c]{
    \resizebox{0.8\textwidth}{!}
    {
    \begin{tabular}{l r r r}
        \hline
        Model &
        \multicolumn{1}{c}{Cu} &
        \multicolumn{1}{c}{$\gamma$-Al$_2$O$_3$} &
        \multicolumn{1}{c}{$\alpha$-Al$_2$O$_3$} \\
        \hline
        & F$_\mathrm{RMSE}$ [eV/\AA] $\downarrow$
        & F$_\mathrm{RMSE}$ [eV/\AA] $\downarrow$
        & F$_\mathrm{RMSE}$ [eV/\AA] $\downarrow$ \\
        \hline
        DP-UniAlCu  &         0.049 &         0.060 &         0.055 \\
        MACE-UniAlCu& \textbf{0.040}& \textbf{0.041}& \textbf{0.030}\\
        MACE-OMAT   &         0.051 &         0.135 &         0.115 \\
        MatterSim-1M&         0.077 &         0.175 &         0.136 \\
        MatterSim-5M&         0.079 &         0.146 &         0.118 \\
        DPA3-OMAT   &         0.050 &         0.140 &         0.088 \\
        \hline
    \end{tabular}
    }
    }
\end{table}

\clearpage

\end{document}